\documentclass{amsart}
\usepackage{amsmath} 
\usepackage{amsthm}
\usepackage{amssymb}
\usepackage{graphicx}
\usepackage{enumerate}

\theoremstyle{plain}
\newtheorem{thm}{Theorem}[section]
\newtheorem{prop}[thm]{Proposition}

\newtheorem{lem}[thm]{Lemma}

\theoremstyle{remark}
\newtheorem{rem}[thm]{Remark}

\theoremstyle{definition}
\newtheorem{ex}[thm]{Example}
\newtheorem{defn}[thm]{Definition}

\def\T{\mathbb{T}}
\def\ds{\displaystyle}

\numberwithin{equation}{section}
\numberwithin{figure}{section}
\numberwithin{table}{section}

\begin{document}

\renewcommand{\thefootnote}{\fnsymbol{footnote}}
\footnote[0]{2000 Mathematics Subject Classification 
Primary 62P05; Secondary 91B28.}


\title[\lambdall]{Optimal intertemporal risk allocation\\
applied to insurance pricing}
\author[\lambdall]{Kei Fukuda, Akihiko Inoue and Yumiharu Nakano}
\address{Japan Credit Rating Agency, Ltd. \\
Jiji Press Building \\
5-15-8 Ginza, Chuo-ku \\
Tokyo 104-0061, Japan}
\email{fukuda@jcra.com}
\address{Department of Mathematics \\
Hokkaido University \\
Sapporo 060-0810 \\
Japan}
\email{inoue@math.sci.hokudai.ac.jp}
\address{Japan Science and Technology Agency \\
Center for Research in Advanced Financial Technology \\
Tokyo Institute of Technology \\
Ookayama 152-8852, Japan}
\email{nakano@craft.titech.ac.jp}

\date{The 1st ver.\ November 8, 2007; this version November 22,2007}


\begin{abstract}
We present a general approach to the pricing of 
products in finance and insurance in the multi-period 
setting. 
It is a combination of the utility indifference pricing and 
optimal intertemporal risk allocation. 
We give a characterization of the optimal intertemporal risk 
allocation by a first order condition. 
Applying this result to the exponential utility function, 
we obtain an essentially new type of premium calculation 
method for a popular type of multi-period insurance contract. 
This method is simple and can be easily implemented numerically. 
We see that the results of numerical calculations are well 
coincident with the risk loading level 
determined by traditional practices. 
The results also suggest a possible implied utility approach 
to insurance pricing.
\end{abstract}

\maketitle


\section{Introduction}\label{sec:1}

The insurer of an insurance contract needs to ensure that the 
premium contains a necessary conservative margin --- the 
so called risk loading or safety loading --- to 
put up the risk capital. 
When determining this margin in a multi-period insurance contract, 
the insurer faces two types of risks to evaluate. 
The first one comes from unfavorable fluctuations in the level of 
investment funded by accumulated premiums. 
The second risk comes from the uncertainty of (life) time, i.e., 
the risk of the unfavorable event occurring at an inopportune time, e.g., 
before the funding target is reached. 
It is desirable to determine the margin that reflects 
both types of risks adequately. 
However, there seems to be no theoretically established solution 
to this challenging problem. The main difficulty is 
in the inseparable nature of the two types of risks themselves; 
the insurance contract guarantees a defined payment at an uncertain time of the insured event occurring by uncertain funding.

In this paper, toward a solution to the problem above, 
we present a fairly general approach to the multi-period pricing problem. 
It is a combination of the {\em utility indifference pricing} and 
{\em optimal intertemporal risk allocation}. 
Though both are quite general concepts, their combination 
leads us to an interesting new premium calculation method 
in a multi-period setting. 

The general setting of the utility indifference pricing is as follows: 
we define the {\it indifference price\/} $H(Z)$ of a risk $Z$ by 
\begin{equation*}
 U(w+H(Z)-Z)=U(w), 
\tag{IP}
\end{equation*}
where $U(W)$ denotes the utility of a risk $W$ and 
the constant $w$ is the initial wealth of the seller of $Z$. 
The price $H(Z)$ is the so-called {\itshape selling} indifference 
price: $H(Z)$ is the amount that leaves 
the seller of the risk $Z$ indifferent between selling and being paid for 
$Z$, and neither selling nor being paid for $Z$. 
In mathematical finance, the 
indifference pricing approach is becoming one of the 
major pricing methods in incomplete markets (see, e.g., 
Hodges and Neuberger \cite{HN}, Rouge and El Karoui \cite{RE}, 
Musiela and Zariphopoulou \cite{MZ}, Bielecki et al.\ \cite{BJR}, and 
M{\o}ller and Steffensen \cite{MS}). 
The indifference pricing also fits the pricing of insurance well. 
For example, in the single-period pricing, 
we can show that many known premium principles are obtained by this method. 
The expectation, variance and exponential premium 
principles are among them. 
Thus, the utility indifference pricing approach has the potential advantage of 
pricing products in finance and insurance coherently.

We write $\mathcal{A}(W)$ for the class of admissible 
intertemporal risk allocations $(Y_t)_{t\in\T}$ of $W$ 
over the multi-period interval $\T:=\{1,2,\dots,T\}$ 
(see Definition \ref{defn:2.1} below): 
$(Y_t)_{t\in\T}$ is an essentially bounded adapted process satisfying 
the risk allocation condition
\begin{equation*}
\sum\nolimits_{t\in\T}\tilde Y_t=W\quad\mbox{a.s.},
\tag{RA}
\end{equation*}
where $\tilde Y_t$ denotes the discounted value of $Y_t$. 
In this paper, we adopt the following utility $U(\cdot)$ in (IP):
\begin{equation*}
U(W):=\sup 
\left\{\sum\nolimits_{t\in\T}E[u_t(\tilde Y_{t})]: 
(Y_t)_{t\in\T}\in\mathcal{A}(W)\right\}.
\tag{U}
\end{equation*}
Here $u_t(x)$ is a time-dependent utility function 
describing the intertemporal preferences of an 
economic agent such as an insurance company. 
This definition says that 
if an allocation $(X_t)\in\mathcal{A}(W)$ attains the supremum in (U), 
then the utility of $W$ is based on the choice of $(X_t)$. 
Thus, to precisely investigate $U(\cdot)$, whence $H(\cdot)$, 
we are led to the problem of finding $(X_t)\in\mathcal{A}(W)$ 
that attains the supremum in (U), which we call the 
{\it optimal intertemporal risk allocation} of $W$.

The optimal risk allocation problems date back to the classical work of 
Borch \cite{Bo1,Bo2,Bo3}, where 
Pareto optimality in uncertain circumstances is studied 
extensively, motivated mainly by reinsurance. 
Since then, various types of optimal risk allocation problems 
have been considered by B\"uhlmann \cite{Bu1,Bu2}, 
Gerber \cite{G}, B\"uhlmann and Jewell \cite{BJ}, and many others. 
See also Gerber and Pafumi \cite{GP}, 
Duffie \cite{D}, Dana and Jeanblanc \cite{DJ} and Dana and Scarsini \cite{DS}. 
Recently, many authors consider the problems based on the preferences 
defined by coherent or convex risk measures introduced by 
Artzner et al.\ \cite{ADEH}, Delbaen \cite{De}, and F\"ollmer and Schied 
\cite{FS1} (see also \cite{FS2}). 
See, e.g., Heath and Ku \cite{HK}, Barrieu and El Karoui \cite{BE}, 
Burgert and R\"uschendorf \cite{BR}, Acciaio \cite{A}, and 
Jouini et al.\ \cite{JST}. 

Unlike most of these references where the problems of 
optimal risk allocation among several economic agents are discussed, 
we consider a single agent in the multi-period framework who 
seeks to find the optimal intertemporal allocation of her/his risk. 
As the definition itself suggests, this optimality is closely related 
to Pareto optimality. 
Note, however, that 
classical Pareto optimality is concerned with 
allocations of risk among economic agents in single-period models, 
while the Pareto optimality we consider 
in this paper is concerned with 
intertemporal allocations of the aggregate risk of a single agent 
in the multi-period setting, 
whence it may be called {\itshape time\/} Pareto optimality.

Our key finding about the optimal intertemporal risk allocation 
(Theorem \ref{thm:2.8}) is that an allocation $(X_t)\in\mathcal{A}(W)$ is optimal 
if and only if the following first order condition is 
satisfied:
\begin{equation*}
 (u_t^{\prime}(\tilde X_t))_{t\in\T} \;\;\text{is an 
 $(\mathcal{F}_t)$-martingale}, 
\tag{FO}
\end{equation*}
where $u_t^{\prime}(x):=(du_t/dx)(x)$ and $(\mathcal{F}_t)_{t\in\T}$ 
is the underlying information structure. 
It is perhaps interesting that this first order condition 
involves a martingale property. 
By applying this characterization to the exponential utility, 
we can derive an algorithm to compute 
the optimal intertemporal risk allocation and 
indifference price $H(\cdot)$ for it (Theorem \ref{thm:3.4}). 
We illustrate the usefulness of this algorithm 
by applying it to a 
popular type of multi-period insurance contract, whereby 
obtaining an essentially 
new type of premium calculation method in the multi-period setting 
(Theorem \ref{thm:4.3}). 
This method is simple and can be easily implemented numerically. 
We see that the results of numerical calculations are well 
coincident with the risk loading level 
determined by traditional practices. 
The results also suggest a possible {\it implied utility approach\/} 
to insurance pricing.

In \S 2, we give basic results on the optimal intertemporal 
risk allocation, including its characterization by (FO) and 
its relationship to Pareto optimality. In \S 3, we apply the results in 
\S 2 to the exponential utility function and derive the 
optimal intertemporal risk allocation and indifference price 
for it. 
Section 4 is devoted to the applications of the results in \S 3 
to insurance pricing. We also discuss properties of 
the indifference prices and some results of numerical calculations.


\section{Optimal intertemporal risk allocation}\label{sec:2}

Let $\T:=\{1,2,\dots,T\}$. 
Throughout the paper, we work on a filtered probability space 
$(\Omega,\mathcal{F},(\mathcal{F}_t)_{t\in\{0\}\cup\T},P)$. 
We write 
$L^{\infty}:=L^{\infty}(\Omega,\mathcal{F}_T,P)$ for the space of 
all essentially bounded, real-valued $\mathcal{F}_T$-measurable random 
variables. 
Let $(r_t)_{t\in\T}$ be a spot rate process. 
We assume that the process $(r_t)_{t\in\T}$ is bounded, 
nonnegative and predictable, i.e., 
$r_t$ is bounded, nonnegative and $\mathcal{F}_{t-1}$-measurable for 
all $t\in\T$. 
Let $B_t$ be the price of the riskless bond:
\[
B_0=1,\qquad 
B_t=\prod_{k=1}^t(1+r_k)\quad \mbox{for}\ t=1,\dots,T.
\]
Throughout the paper, 
we use $(B_t)_{t\in\T}$ as the num\'eraire, and for each price process 
$(X_t)_{t\in\T}$, 
we denote by $(\tilde X_t)_{t\in\T}$ its 
discounted price process:
\[
\tilde X_t:=X_t/B_t,\qquad t\in\T.
\]

\subsection{Optimality}\label{subsec:2.1} 

We consider an economic agent such as an insurance company 
who wishes to allocate her/his aggregate risk $W$ over the 
multi-period interval $\T$. 
In the next definition, we define the collection of all such possible 
intertemporal allocations of $W$.

\begin{defn}\label{defn:2.1}
For $W\in L^{\infty}$, we write 
$\mathcal{A}(W)$ for the following set of {\itshape admissible intertemporal 
allocations\/} $(Y_t)_{t\in\T}$ of $W$: 
\[
\mathcal{A}(W):=\left\{(Y_t)_{t\in\T}: 
\begin{aligned}
&\mbox{$(Y_t)_{t\in\T}$ is an $(\mathcal{F}_t)$-adapted process 
satisfying}\\
&\mbox{(RA) and $Y_t\in L^{\infty}$ for all $t\in\T$.}
\end{aligned}
\right\}.
\]
\end{defn}

\begin{ex}\label{ex:2.2}
We consider the aggregate risk $W$ of a life insurance contract with 
duration $T$ 
in which the insured receives $c_t$ dollars at time $t\in\T$ if she/he dies 
in the period $(t-1,t]$. Then, we have $W=\sum_{t\in\T}\tilde Y_t$ with 
$Y_t:=c(t)1_{(t-1<\tau\le t)}$, where $\tau$ is the stopping time representing 
the lifetime of the insured. 
Notice that $(Y_t)_{t\in\T}$ itself is in 
$\mathcal{A}(W)$. 
If we define $(X_t)_{t\in\T}$ by
\begin{equation*}
 X_t=\begin{cases}
      0, & \qquad t=1, \\
      (1+r_t)Y_{t-1}, & \qquad t=2,\dots,T-1, \\
      Y_T+(1+r_T)Y_{T-1}, & \qquad t=T,
     \end{cases}
\end{equation*}
then $(X_t)_{t\in\T}$ is also in $\mathcal{A}(W)$. 
Insurance companies which have many contracts with policyholders 
will be able to regard $W$ 
as the aggregate risk of $(X_t)_{t\in\T}$, rather than that of 
$(Y_t)_{t\in\T}$, 
at a negligible cost.
\end{ex}

We assume that the intertemporal preferences of the agent is described by 
the time-dependent utility function $u_t(x)$. 
This means that a rational choice of the agent's 
allocation $(Y_t)_{t\in\T}\in\mathcal{A}(W)$ is based on the integrated 
expected utility $\sum_{t\in\T}E[u_t(\tilde Y_t)]$. 
Throughout \S \ref{sec:2}, we assume that the utility function $u_t(x)$ 
satisfies the following condition: 
\begin{equation}
\begin{cases}
\mbox{for $t\in\T$, 
$\mathbb{R}\ni x\mapsto u_t(x)\in \mathbb{R}$ is a 
strictly concave, $C^1$-class function}\\
\mbox{such that 
$u_t'(x):=(du_t/dx)(x)>0$ for $x\in\mathbb{R}$.}
\end{cases}
\label{eq:2.1}
\end{equation}
Using $u_t(x)$, 
we define the utility $U(W)\in \mathbb{R}\cup\{+\infty\}$ of the risk 
$W\in L^{\infty}$ by (U).

\begin{defn}\label{defn:2.3}
An intertemporal risk allocation $(X_t)_{t\in\T}\in\mathcal{A}(W)$ of 
the risk $W\in L^{\infty}$ is {\it optimal\/} if it attains 
the supremum in (U).
\end{defn}

In other words, $(X_t)_{t\in\T}\in\mathcal{A}(W)$ is optimal if it 
solves the following problem:
\begin{equation*}
\mbox{Maximize}\quad 
\sum\nolimits_{t\in\T}E[u_t(\tilde Y_t)]\quad \mbox{among all}\quad 
(Y_t)_{t\in\T}\in\mathcal{A}(W).
\tag{P}
\end{equation*}

\begin{prop}\label{prop:2.4}
The optimal intertemporal risk allocation 
$(X_t)_{t\in\T}\in\mathcal{A}(W)$ of $W\in L^{\infty}$ 
is unique if it exists.
\end{prop}

\begin{proof}
Suppose that there are two distinct optimal 
intertemporal allocations $(X_t)$ and 
$(Y_t)$ of $W$. 
If we put $Z_t:=(1/2)X_t+(1/2)Y_t$ for $t\in\T$, then 
$(Z_t)$ is also in $\mathcal{A}(W)$. However, concavity of 
$u_t(\cdot)$ yields
\begin{equation*}
 \sum\nolimits_{t\in\T} E[u_t( \tilde Z_t)]
 >\sum\nolimits_{t\in\T} E[(1/2)u_t( \tilde X_t)
 +(1/2)u_t(\tilde Y_t)]=U(W),
\end{equation*}
which is a contradiction. Thus the optimal allocation of $W$ is unique.
\end{proof}

\subsection{Indifference pricing}\label{subsec:2.2}

In this section, we assume that $U(W)<\infty$ for all $W\in L^{\infty}$. 
This condition holds, for example, if $u_t(x)$ is bounded from above. 
This also holds if the optimal intertemporal risk allocation exists for 
all $W\in L^{\infty}$. 
We thus have the utility functional $U:L^{\infty}\to \mathbb{R}$. 
We write $w\in \mathbb{R}$ for the initial wealth of the agent.

\begin{prop}\label{prop:2.5}
The functional $U$ has the following properties for 
$W, Z\in L^{\infty}$.
\begin{itemize}
\item[{\rm (a)}] Strict Monotonicity: If $W\ge Z$ a.s.\ and $P(W>Z)>0$, 
then $U(W)>U(Z)$.
\item[{\rm (b)}] Concavity: If $a\in [0,1]$, then 
$U(aW + (1-a)Z) \ge a U(W) + (1-a) U(Z)$.
\end{itemize}
\end{prop}

\begin{proof}
(a)\ 
For $(Y_t)_{t\in\T}\in\mathcal{A}(Z)$, 
we define $(X_{t})_{t\in\T}\in\mathcal{A}(W)$ by
\begin{equation*}
X_t=
\begin{cases}
Y_t, & t\neq T, \\
Y_T+ B_T(W-Z), & t=T.
\end{cases}
\end{equation*}
Choosing $m>0$ so that $\max(\vert W\vert, \vert Z\vert)\le m$ a.s., we 
define $c:=\inf_{\vert y\vert\le m}u_T'(y)$. Then, by (\ref{eq:2.1}), 
$c>0$. 
Since $u_T(\tilde X_T)\ge u_T(\tilde Y_T) + c(W-Z)$, we have
\[
U(W)\ge \sum\nolimits_{t\in\T}E[u_t(\tilde X_t)]
\ge \sum\nolimits_{t\in\T}E[u_t(\tilde Y_t)]
+cE[W-Z].
\]
The property (a) follows from this.

(b)\ The property (b) follows easily from the 
concavity of $u_t$, $t\in\T$.
\end{proof}

From Proposition \ref{prop:2.5}, we see that 
for $Z \in L^{\infty}$, 
the function $g: \mathbb{R}\to\mathbb{R}$ defined by $g(x):=U(w+x-Z)$ 
is concave (whence continuous) and strictly increasing. 
Moreover, since $Z$ is bounded, we have $U(w+x-Z)<U(w)$ for $x$ small 
enough and 
$U(w+x-Z)>U(w)$ for $x$ large enough. 
We are thus led to the following definition.

\begin{defn}\label{defn:2.6}
We define the {\it indifference price\/} 
$H(Z)=H(Z;w)\in\mathbb{R}$ 
of $Z\in L^{\infty}$ by 
$U(w+H(Z) - Z)=U(w)$.
\end{defn}

From Proposition \ref{prop:2.5}, we immediately obtain the next 
proposition.

\begin{prop}\label{prop:2.7}
The indifference price functional 
$H: L^{\infty}\to \mathbb{R}$ has the following 
propertites for $W, Z\in L^{\infty}$.
\begin{itemize}
\item[{\rm (a)}] Strict Monotonicity: If $W\ge Z$ a.s.\ and $P(W>Z)>0$, 
then $H(W)>H(Z)$.
\item[{\rm (b)}] Convexity: If $a\in [0,1]$, then 
$H(aW + (1-a)Z) \le a H(W) + (1-a) H(Z)$.
\end{itemize}
\end{prop}

\subsection{Characterization by the first order condition}\label{subsec:2.3}

It should be noticed that, in general, the optimal intertemporal 
risk allocation may not exist. 
However, to precisely investigate the utility $U(\cdot)$, whence 
the indifference price $H(\cdot)$, it seems indispensable 
to find and describe the optimal intertemporal risk allocation. 
In this section, we show that the condition (FO) is necessary and sufficient 
for $(X_t)\in\mathcal{A}(W)$ to be optimal. 
This characterization plays a key role in this paper. 
In the proof below, and throughout the paper, we write
\[
E_t[Y]:=E[Y\vert \mathcal{F}_t],\qquad 
Y\in L^1(\Omega,\mathcal{F},P),\ t\in\T.
\]

Here is the characterization of the optimality.

\begin{thm}\label{thm:2.8}
For $W\in L^{\infty}$ and $(X_t)_{t\in\T}\in\mathcal{A}(W)$, 
the following conditions are equivalent:
\begin{itemize}
\item[{\rm (a)}]
$(X_t)_{t\in\T}$ is optimal.
\item[{\rm (b)}]
The condition {\rm (FO)} is satisfied.
\end{itemize}
\end{thm}

\begin{proof}
First, we prove (a) $\Rightarrow$ (b). 
Let $(X_t)\in \mathcal{A}(W)$ be the optimal allocation. 
Choose $k,m\in\T$ so that $k<m$, and put, 
for $t\in\T$, $y\in\mathbb{R}$ and $A\in\mathcal{F}_k$,
\begin{equation*}
 X_t(y)=\begin{cases}
      X_{m} + yB_{m}1_A, & \qquad t=m, \\
      X_k - yB_k1_A, & \qquad t=k, \\
      X_t, & \qquad\mbox{otherwise}.
     \end{cases}
\end{equation*}
Then, $\sum_{t\in\T}\tilde X_t(y)=W$, 
so that $(X_t(y))_{t\in\T}\in\mathcal{A}(W)$. 
Since $(X_t(0))=(X_t)$ is optimal, 
the function $f$ defined by 
$f(y):=\sum\nolimits_{t\in\T} E[u_t(\tilde X_t(y))]$ 
takes the maximal value at $y=0$. Thus 
$f^{\prime}(0)=0$ or 
$E[\{u_m'(\tilde X_{m})-u_k'(\tilde X_k)\}1_A]=0$, 
which implies that $(u_t'( \tilde X_t))$ is an 
$(\mathcal{F}_t)$-martingale.

Next, we prove (b) $\Rightarrow$ (a). Assume that 
$(X_t)_{t\in\T}\in\mathcal{A}(W)$ and that 
$(u_t'(\tilde X_t))_{t\in\T}$ is an $(\mathcal{F}_t)$-martingale. 
By concavity of 
$u_t(\cdot)$, we have 
$u_t(y)\le u_t(x) + u_t'(x)(y-x)$ for $x,y\in\mathbb{R}$, 
so that for any $Y=(Y_t)_{t\in\T}\in\mathcal{A}(W)$, 
\begin{equation*}
\sum\nolimits_{t\in\T} u_t(\tilde Y_t)
\le\sum\nolimits_{t\in\T}
 u_t(\tilde X_t)
  +\sum\nolimits_{t\in\T} u_t'(\tilde X_t)(\tilde Y_t-\tilde X_t).
\end{equation*}
Since $(u_t'(\tilde X_t))$ is an 
$(\mathcal{F}_t)$-martingale and 
both $(X_t)$ and $(Y_t)$ are in $\mathcal{A}(W)$, 
we see that
\begin{align*}
&E\left[\sum\nolimits_{t\in\T} u_t'(\tilde X_t)
(\tilde Y_t-\tilde X_t)\right]\\
&=\sum\nolimits_{t\in\T} E\left[E_t[u_T'( \tilde X_{T})]
(\tilde Y_t-\tilde X_t)\right]
=\sum\nolimits_{t\in\T} E\left[u_T'( \tilde X_{T})
(\tilde Y_t-\tilde X_t)\right]\\
&=E\left[u_T'(\tilde X_{T})
\sum\nolimits_{t\in\T}(\tilde Y_t-\tilde X_t)\right]
=E\left[u_T'(\tilde X_{T})(W-W)\right]
=0.
\end{align*}
Combining, 
$\sum\nolimits_{t\in\T} E[u_t(\tilde Y_t)]
\le \sum\nolimits_{t\in\T} E[u_t( \tilde X_t)]$. Thus, 
$(X_t)$ is optimal.
\end{proof}

\begin{rem}\label{rem:2.9}
We clearly find similarity between the theorem above and 
Borch's theorem which characterizes (classical) Pareto optimality by a 
first order 
condition (see Borch \cite{Bo1,Bo2,Bo3}; 
see also Gerber and Pafumi \cite{GP}).
\end{rem}

\subsection{Pareto optima}\label{subsec:2.4}

In this section, we introduce Pareto optimality of intertemporal 
risk allocations. It is closely related to the optimality 
introduced above.

\begin{defn}\label{defn:2.10}
For $W\in L^{\infty}$, the allocation 
$(X_t)_{t\in\T}\in\mathcal{A}(W)$ 
is {\itshape Pareto optimal\/} if there does not exist 
$(Y_t)_{t\in\T}\in \mathcal{A}(W)$ 
satisfying the following two conditions:
\begin{itemize}
\item[(a)] 
$E[u_t(\tilde Y_t)]\ge E[u_t(\tilde X_t)]$ for all $t\in\T$.
\item[(b)] 
$E[u_{t_0}(\tilde Y_{t_0})]> E[u_{t_0}(\tilde X_{t_0})]$ 
for at least one $t_0\in \T$.
\end{itemize}
\end{defn}

For $\lambda=(\lambda_1,\dots,\lambda_T)\in \mathbb{R}_+^{T}\setminus\{0\}$, 
we consider the following problem:
\begin{equation*}
\mbox{Maximize}\quad 
\sum\nolimits_{t\in\T}\lambda_tE[u_t(\tilde Y_t)]\quad \mbox{among all}\quad 
(Y_t)_{t\in\T}\in\mathcal{A}(W).
\tag{$\mathrm{P}_{\lambda}$}
\end{equation*}

\begin{lem}\label{lem:2.11}
Let $\lambda=(\lambda_1,\dots,\lambda_T)\in\mathbb{R}_+^{T}\setminus\{0\}$. 
\begin{itemize}
\item[{\rm (a)}]If $(X_t)_{t\in\T}\in\mathcal{A}(W)$ is the solution to 
Problem $\mathrm{P}_{\lambda}$, then $(\lambda_tu_t'(\tilde X_t))_{t\in\T}$ 
is an $(\mathcal{F}_t)$-martingale.
\item[{\rm (b)}]If Problem $\mathrm{P}_{\lambda}$ has a solution, then 
$\lambda\in (0,\infty)^{T}$.
\end{itemize}
\end{lem}

\begin{proof}
The proof of (a) is almost 
the same as that of the implication (a) $\Rightarrow$ (b) 
in Theorem \ref{thm:2.8}, whence we omit it.

We prove (b). Assume that $\lambda_k=0$ for $k\in\T$, and 
choose $m$ so that $\lambda_{m}>0$. 
If Problem $\mathrm{P}_{\lambda}$ has a solution $(X_t)\in\mathcal{A}(W)$, 
then, by (a), $(\lambda_tu_t'(\tilde X_t))$ is an 
$(\mathcal{F}_t)$-martingale. 
However, 
since $\lambda_ku_k'(\tilde X_k)=0$ and $\lambda_mu_m'(\tilde X_m)>0$, 
this can never be the case. Thus, (b) follows.
\end{proof}

\begin{prop}\label{prop:2.12}
Let $\lambda\in (0,\infty)^{T}$. Then the solution 
$(X_t)_{t\in\T}\in\mathcal{A}(W)$ to 
Problem $\mathrm{P}_{\lambda}$ is unique if exists.
\end{prop}

The proof is almost the same as that of Proposition \ref{prop:2.4}, 
whence we omit it.

The next theorem is an analogue of the 
{\itshape second fundamental theorem of welfare economics}.

\begin{thm}\label{thm:2.13}
For $(X_t)_{t\in\T}\in\mathcal{A}(W)$, the following conditions are 
equivalent:
\begin{itemize}
\item[{\rm (a)}] $(X_t)_{t\in\T}$ is Pareto optimal.
\item[{\rm (b)}] There exists $\lambda\in (0,\infty)^{T}$ such that 
$(X_t)_{t\in\T}$ solves Problem $\mathrm{P}_{\lambda}$.
\end{itemize}
\end{thm}

\begin{proof}
\noindent (b) $\Rightarrow$ (a). 
If $(X_t)_{t\in\T}$ is not Pareto optimal, then clearly it is not 
the solution to Problem $\mathrm{P}_{\lambda}$ for any 
$\lambda\in (0,\infty)^{T}$. 

\noindent (a) $\Rightarrow$ (b). 
We define $f(Y):=\phi(X)-\phi(Y)$ for $Y\in\mathcal{A}(W)$, where
\[
\phi(Y):=\left(E[u_1(\tilde Y_1)],\dots,E[u_T(\tilde Y_T)]\right).
\]
Then $f:\mathcal{A}(W)\to \mathbb{R}^{T}$ is $\mathbb{R}_{+}^{T}$-convex: 
for $p\in (0,1)$ and $Y, Y'\in\mathcal{A}(W)$,
\[
p f(Y) + (1-p)f(Y') - f(p Y + (1-p)Y')
\in \mathbb{R}_{+}^{T}.
\]
If $X\in\mathcal{A}(W)$ is Pareto optimal, then 
$-f(Y)\notin (0,\infty)^T$ for 
$Y\in\mathcal{A}(W)$. 
Hence, by Gordan's Alternative Theorem 
(see, e.g., Craven \cite{C}, Chapter 2), there exists 
$\lambda\in \mathbb{R}_{+}^{T}$, $\lambda\ne 0$, such that
\begin{equation*}
\lambda\cdot f(Y)=\lambda\cdot\left[\phi(X)-\phi(Y)\right]\ge 0, 
 \qquad Y\in\mathcal{A}(W),
\end{equation*}
which implies that $X$ is the solution 
to Problem $\mathrm{P}_{\lambda}$. 
Finally, Lemma \ref{lem:2.11} gives 
$\lambda\in (0,\infty)^{T}$.
\end{proof}

By Theorem \ref{thm:2.13}, we see that the set of Pareto optimal 
intertemporal risk allocations in $\mathcal{A}(W)$ is parametrized by 
the $T-1$ parameters $(\lambda_2/\lambda_1,\dots,\lambda_T/\lambda_1)\in 
(0,\infty)^{T-1}$. We also see that 
the Pareto optimal allocation $(X_t)\in\mathcal{A}(W)$ 
corresponding to Problem ($\mathrm{P}_{\lambda}$) with 
$\lambda=(\lambda_1,\dots,\lambda_T)$ is optimal with respect to 
the intertemporal preferences described by the utility 
function $v_t(x):=\lambda_tu_t(x)$. 
Therefore, from Theorem \ref{thm:2.8}, we immediately obtain the next 
characterization of Pareto optimality.

\begin{thm}\label{thm:2.14}
For $W\in L^{\infty}$ and $(X_t)_{t\in\T}\in\mathcal{A}(W)$, 
the following conditions are equivalent:
\begin{itemize}
\item[{\rm (a)}]
$(X_t)_{t\in\T}$ is Pareto optimal.
\item[{\rm (b)}]
There exists $(\lambda_1,\dots,\lambda_T)\in (0,\infty)^{T}$ such that 
the process $(\lambda_tu_t^{\prime}(\tilde X_t))_{t\in\T}$ is an 
$(\mathcal{F}_t)$-martingale.
\end{itemize}
\end{thm}


\section{Exponential utility}\label{sec:3}

Let $(r_t)_{t\in\T}$ and $(B_t)_{t\in\T}$ be as in Section \ref{sec:2}. 
In this section, we adopt the following time-dependent exponential utility 
function:
\begin{equation*}
\left\{
\begin{aligned}
&u_t(x)=\frac{1}{\alpha_{t}}\left[1-\exp\left(-\alpha_{t}x\right)\right],
\qquad t\in\T,\ x\in\mathbb{R}\\
&\mbox{with $\alpha:=(\alpha_1,\dots,\alpha_{t})\in (0,\infty)^T$.}
\end{aligned}
\right.
\tag{EU}
\end{equation*}
In what follows, we may also write
$\alpha(t)=\alpha_t$.
We have
\begin{equation}
u_t'(x)=\exp\left(-\alpha_{t}x\right),\qquad 
u_t(0)=0,\qquad u_t'(0)=1.
\label{eq:3.1}
\end{equation}

\subsection{The optimal allocation for the 
exponential utility}\label{subsec:3.1}

In this section, we describe the optimal intertemporal risk allocation 
for the exponential utility function $u_t(x)$ in (EU). 
Thus, the problem that we consider here is Problem (P) 
for $u_t(x)$ in (EU).

To derive the optimal allocation $(X_t)_{t\in\T}\in\mathcal{A}(W)$ or 
the solution to (P), 
we consider the transform 
$M_t=\exp(-\alpha_{t}\tilde X_t)$ for $t\in\T$. 
Then, by Theorem \ref{thm:2.8}, Problem (P) reduces to 

\ 

\noindent {\bf Problem M}.\quad 
For $W\in L^{\infty}$ and 
$\alpha=(\alpha_1,\dots,\alpha_T)\in (0,\infty)^T$, 
derive a positive $(\mathcal{F}_t)$-martingale $(M_t)_{t\in\T}$ satisfying
\begin{equation}
\prod\nolimits_{t\in\T}M_t^{1/\alpha(t)}=\exp(-W)\qquad\mbox{a.s.}
\label{eq:3.2}
\end{equation}

For $W\in L^{\infty}$ and 
$\alpha=(\alpha_1,\dots,\alpha_T)\in (0,\infty)^T$, we define 
the adapted process $(L_t(\alpha, W))_{t\in\T}$ by the following 
backward iteration:
\begin{equation*}
\begin{cases}
L_T(\alpha, W):=\exp(-\alpha_{T}W), & \\
L_{t-1}(\alpha, W):
=E_{t-1}[L_t(\alpha, W)]^{\beta(t-1)/\beta(t)}, & 
t=2,\dots,T,
\end{cases}
\tag{L1}
\end{equation*}
where $E_t[Y]:=E[Y \vert \mathcal{F}_t]$ as before, and 
we define $\beta_t$, or $\beta(t)$, in $(0,\infty)$ by
\begin{equation*}
\frac{1}{\beta_{t}}=\sum_{k=t}^{T}\frac{1}{\alpha_{k}},\qquad 
t\in\T.
\tag{$\beta$}
\end{equation*}
Notice that for all $t\in\T$, $L_t(\alpha, W)$ is bounded away from $0$ 
and $\infty$. 
We also define the adapted 
process $(M_t(\alpha, W))_{t\in\T}$ by
\begin{equation*}
\begin{cases}
M_t(\alpha, W)=L_t(\alpha, W)\cdot\prod_{k=1}^{t-1}
L_k(\alpha, W)^{-\beta(k+1)/\alpha(k)}, & t=2,\dots,T,\\
M_1(\alpha, W)=L_1(\alpha, W). & 
\end{cases}
\tag{M}
\end{equation*}

Here is the solution to the martingale problem $M$ above.

\begin{thm}\label{thm:3.1}
For $W\in L^{\infty}$ and 
$\alpha=(\alpha_1,\dots,\alpha_T)\in (0,\infty)^T$, 
the solution $(M_t)_{t\in\T}$ 
to Problem M is unique and given by $M_t=M_t(\alpha, W)$ for $t\in\T$.
\end{thm}

\begin{proof}
For simplicity, we write $L_t=L_t(\alpha,W)$ for $t\in\T$.

\noindent {\itshape Step 1.}\quad Let $t\ge 3$. 
Since $\prod_{k=1}^{t-1}L_k^{-\beta(k+1)/\alpha(k)}$ is 
$\mathcal{F}_{t-1}$-measurable, 
the process $(M_t)_{t\in\T}$ defined by $M_t=M_t(\alpha, W)$ satisfies
\[
E_{t-1}[M_t]
=E_{t-1}[L_t]\cdot\prod\nolimits_{k=1}^{t-1}
L_k^{-\beta(k+1)/\alpha(k)}.
\]
However, since $E_{t-1}[L_t]=L_{t-1}^{\beta(t)/\beta(t-1)}$, 
we get
\begin{align*}
E_{t-1}[M_t]
&=
L_{t-1}^{\beta(t)/\beta(t-1)}\cdot 
L_{t-1}^{-\beta(t)/\alpha(t-1)}\cdot \prod\nolimits_{k=1}^{t-2}
L_k^{-\beta(k+1)/\alpha(k)}\\
&=L_{t-1}\cdot \prod\nolimits_{k=1}^{t-2}
L_k^{-\beta(k+1)/\alpha(k)}=M_t.
\end{align*}
Treating the case $t=2$ similarly, we see that 
$(M_t)$ is an $(\mathcal{F}_t)$-martingale. Also,
\begin{align*}
\prod_{t\in\T} M_t^{1/\alpha(t)}
&=L_1^{1/\alpha(1)}\cdot\prod\nolimits_{t=2}^TL_t^{1/\alpha(t)}
\left(\prod\nolimits_{k=1}^{t-1}
L_k^{-\beta(k+1)/\{\alpha(k)\alpha(t)\}}\right)\\
&=\left[\prod\nolimits_{t\in\T} L_t^{1/\alpha(t)}\right]
\cdot \left[\prod\nolimits_{t=2}^T\prod\nolimits_{k=2}^{t}
L_{k-1}^{-\beta(k)/\{\alpha(k-1)\alpha(t)\}}\right]\\
&=\left[\prod\nolimits_{t\in\T} L_t^{1/\alpha(t)}\right]
\cdot \left[
\prod\nolimits_{k=2}^{T}
\left(
\prod\nolimits_{t=k}^T
L_{k-1}^{-1/\alpha(t)}
\right)^{-\beta(k)/\alpha(k-1)}\right]\\
&=\left[\prod\nolimits_{t\in\T} L_t^{1/\alpha(t)}\right]
\cdot \left[
\prod\nolimits_{k=2}^{T}
L_{k-1}^{-1/\alpha(k-1)}
\right]=L_T^{1/\alpha(T)},
\end{align*}
yielding (\ref{eq:3.2}). Thus $(M_t)$ is a solution to Problem M.

\noindent {\itshape Step 2.}\quad 
We show the uniqueness. 
Assume that $(M_t)_{t\in\T}$ is a solution to Problem M. 
Then, 
\begin{equation}
\left[\prod\nolimits_{k=1}^{T-2}M_k^{1/\alpha(k)}\right]
\cdot M_{T-1}^{1/\beta(T-1)}
=E_{T-1}[L_T]^{1/\alpha(T)}.
\label{eq:3.3}
\end{equation}
From this, we have the decomposition
\begin{equation}
M_{T-1}=L_{T-1}\cdot N_{T-2},
\label{eq:3.4}
\end{equation}
where $N_{T-2}$ is an $\mathcal{F}_{T-2}$-measurable random variable. 
We see that $N_{T-2}$ satisfies
\[
\left[\prod\nolimits_{k=1}^{T-2}M_k^{1/\alpha(k)}\right]
\cdot N_{T-2}^{1/\beta(T-1)}=1.
\]
However,
\begin{equation*}
M_{T-2}
=E_{T-2}[M_{T-1}]
=E_{T-2}[L_{T-1}]\cdot N_{T-2}\\
=L_{T-2}^{\beta(T-1)/\beta(T-2)}\cdot N_{T-2},
\end{equation*}
so that
\begin{equation*}
\left[\prod\nolimits_{k=1}^{T-3}M_k^{1/\alpha(k)}\right]
\cdot 
N_{T-2}^{1/\beta(T-2)}
=L_{T-2}^{-\beta(T-1)/\{\alpha(T-2)\beta(T-2)\}}.
\end{equation*}
Thus, $N_{T-2}$ also has the decomposition
\[
N_{T-2}=L_{T-2}^{-\beta(T-1)/\alpha(T-2)}\cdot N_{T-3},
\]
where $N_{T-3}$ is $\mathcal{F}_{T-3}$-measurable. 
Moreover, this and (\ref{eq:3.4}) give
\begin{equation}
M_{T-1}=L_{T-1}\cdot L_{T-2}^{-\beta(T-1)/\alpha(T-2)}
\cdot N_{T-3}.
\label{eq:3.5}
\end{equation}

The random variable $N_{T-3}$ satisfies
\[
\left[\prod\nolimits_{k=1}^{T-3}M_k^{1/\alpha(k)}\right]
\cdot N_{T-3}^{1/\beta(T-2)}=1.
\]
However, from
\begin{align*}
&E_{T-2}[L_{T-1}]=L_{T-2}^{\beta(T-1)/\beta(T-2)},\\
&E_{T-3}[L_{T-2}]=L_{T-3}^{\beta(T-2)/\beta(T-3)},
\end{align*}
we find that
\begin{align*}
M_{T-3}
&=E_{T-3}[M_{T-1}]=E_{T-3}[L_{T-1}\cdot L_{T-2}^{-\beta(T-1)/\alpha(T-2)}]
\cdot N_{T-3}\\
&=E_{T-3}[E_{T-2}[L_{T-1}]\cdot L_{T-2}^{-\beta(T-1)/\alpha(T-2)}]
\cdot N_{T-3}\\
&=E_{T-3}[L_{T-2}]\cdot N_{T-3}
=L_{T-3}^{\beta(T-2)/\beta(T-3)}\cdot N_{T-3}.
\end{align*}
Therefore,
\begin{equation*}
\left[\prod\nolimits_{k=1}^{T-4}M_k^{1/\alpha(k)}\right]
\cdot 
N_{T-3}^{1/\beta(T-3)}
=L_{T-3}^{-\beta(T-2)/\{\alpha(T-3)\beta(T-3)\}},
\end{equation*}
so that $N_{T-3}$ has the decomposition
\[
N_{T-3}=L_{T-3}^{-\beta(T-2)/\alpha(T-3)}\cdot N_{T-4},
\]
where $N_{T-4}$ is $\mathcal{F}_{T-4}$-measurable. 
Moreover, from this and (\ref{eq:3.5}), we get
\begin{equation*}
M_{T-1}
=L_{T-1}\cdot L_{T-2}^{-\beta(T-1)/\alpha(T-2)}\cdot 
L_{T-3}^{-\beta(T-2)/\alpha(T-3)}\cdot 
N_{T-4}.
\end{equation*}

Repeating the arguments above, we finally obtain
\[
M_{T-1}=L_{T-1}\cdot \prod\nolimits_{k=1}^{T-2}
L_k^{-\beta(k+1)/\alpha(k)}.
\]
On the other hand, 
we find from (\ref{eq:3.2}) and (\ref{eq:3.3}) that
\[
M_T=\frac{M_{T-1}\cdot L_T}{E_{T-1}[L_T]}.
\]
Moreover, 
$E_{T-1}[L_T]=L_{T-1}^{\alpha(T)/\beta(T-1)}$. 
Combining,
\begin{align*}
M_T
&=L_T\cdot L_{T-1}^{-\alpha(T)/\beta(T-1)}\cdot M_{T-1}\\
&=L_T\cdot L_{T-1}^{-\alpha(T)/\beta(T-1)}\cdot L_{T-1}\cdot 
\prod\nolimits_{k=1}^{T-2}
L_k^{-\beta(k+1)/\alpha(k)}\\
&=L_T\cdot \prod\nolimits_{k=1}^{T-1}
L_k^{-\beta(k+1)/\alpha(k)}.
\end{align*}
Thus $M_T$ coincides with $M_T(\alpha, W)$. 
However, since both $(M_t)$ and $(M_t(\alpha,W))$ are 
$(\mathcal{F}_t)$-martingales, this implies that the two 
processes are identical. 
Thus the solution to Problem M is unique.
\end{proof}

The next theorem follows immediately 
from Theorems \ref{thm:2.8} and \ref{thm:3.1}.

\begin{thm}\label{thm:3.2}
The optimal intertemporal risk allocation 
$(X_t)_{t\in\T}\in\mathcal{A}(W)$ of $W\in L^{\infty}$ for 
the exponential utility function $u_t(x)$ in {\rm (EU)\/} 
is unique and given by
\begin{equation}
\exp(-\alpha_{t}\tilde X_t)=M_t(\alpha,W),\qquad t\in\T.
\label{eq:3.6}
\end{equation}
\end{thm}

We need the next proposition later.

\begin{prop}\label{prop:3.3}
Let $x\in\mathbb{R}$, $Z\in L^{\infty}$ and 
$\alpha=(\alpha_1,\dots,\alpha_T)\in (0,\infty)^T$. Then, 
the following assertions hold:
\begin{itemize}
\item[(a)] 
$L_t(\alpha, x)=\exp(-\beta_{t}x)$ for $t\in\T$.
\item[(b)] 
$L_t(\alpha, x-Z)=\exp(-\beta_{t}x)L_t(\alpha,-Z)$ for $t\in\T$.
\end{itemize}
\end{prop}

\begin{proof}
The assertion (a) follows immediately from the definition of 
$(L_t(\alpha, x))$. 
If we put $L'_t:=\exp(-\beta_{t}x)L_t(\alpha, -Z)$ 
for $t\in\T$, 
then $(L'_t)_{t\in\T}$ satisfies
\begin{equation*}
\begin{cases}
L'_T=\exp\left[-\alpha_{T}(x-Z)\right],&\\
L'_{t-1}
=E_{t-1}[L'_t]^{\beta(t-1)/\beta(t)},
& t=2,\dots,T,
\end{cases}
\label{eq:L-dash}
\end{equation*}
whence $L'_t=L_t(\alpha, x-Z)$ for $t\in\T$ or (b).
\end{proof}


\subsection{The indifference prices for the exponential utility}
\label{subsec:3.2}

In this section, we derive the indifference prices for the exponential 
utility $u_t(x)$ in (EU). 
Let $U, H: L^{\infty}\to\mathbb{R}$ be the 
utility and indifference price functionals 
defined from $u_t(x)$ as above, respectively. 
Recall $\beta_{t}$, $L_t(\alpha,Z)$ and $M_t(\alpha, Z)$ from 
Section \ref{subsec:3.1}.

For the exponential utility, 
the next theorem reduces the computation of the indifference price 
$H(Z)$ to that of $L_1(\alpha, -Z)$.

\begin{thm}\label{thm:3.4}
We assume $({\rm EU})$. Then, 
for $x\in\mathbb{R}$ and $Z\in L^{\infty}$, 
the following assertions hold:
\begin{itemize}
\item[(a)] $\ds U(Z)=\frac{1}{\beta_{1}}\{1-E[L_1(\alpha, Z)]\}$.
\item[(b)] $\ds U(x-Z)
=\frac{1}{\beta_{1}}\{1-\exp(-\beta_{1}x)\cdot E[L_1(\alpha, -Z)]\}$.
\item[(c)] $\ds H(Z)=\frac{1}{\beta_{1}}\log E[L_1(\alpha, -Z)]$.
\end{itemize}
\end{thm}

\begin{proof}
Define $(X_t)_{t\in\T}\in\mathcal{A}(Z)$ by (\ref{eq:3.6}) with $W=Z$.  
Then, by Theorem \ref{thm:3.2}, the supremum in (U) 
is attained by $(X_t)$. Since $(M_t(\alpha, Z))_{t\in\T}$ is an 
$(\mathcal{F}_t)$-martingale and 
$M_1(\alpha, Z)=L_1(\alpha, Z)$, we have
\begin{align*}
U(Z)
&=\sum\nolimits_{t\in\T}\frac{1}{\alpha_{t}}E[1-\exp(-\alpha_{t}\tilde X_t)]
=\sum\nolimits_{t\in\T}\frac{1}{\alpha_{t}}E[1-M_t(\alpha, Z)]\\
&=\{1-E[M_1(\alpha, Z)]\}\sum\nolimits_{t\in\T}\frac{1}{\alpha_{t}}
=\frac{1}{\beta_{1}}\{1-E[L_1(\alpha, Z)]\}.
\end{align*}
Thus (a) follows. The assertion (b) follows from (a) and 
Proposition \ref{prop:3.3} (b). Finally, (c) follows from 
(a), (b) and Proposition \ref{prop:3.3} (a).
\end{proof}

From Theorem \ref{thm:3.4} (c), we see that the indifference price 
$H(Z)$ does not depend on the level $w$ of the initial wealth 
for the exponential utility function.

The next proposition describes the optimal intertemporal allocation 
of the selling position $w+H(Z)-Z$ for the exponential utility.

\begin{prop}\label{prop:3.5}
We assume $({\rm EU})$. 
For $x\in\mathbb{R}$ and $Z\in L^{\infty}$, 
let $(X_t)\in\mathcal{A}(x - Z)$ be the optimal intertemporal 
allocation of $x - Z$: 
$\sum_{t\in\T}E[u_t(\tilde X_{t})] = U(x - Z)$. 
Then, $(X_t)_{t\in\T}$ is given by
\begin{align*}
&X_1=
\frac{B_1}{\alpha_{1}}
\left[
\beta_{1}x - 
\log L_1(\alpha, -Z)
\right], \\
&X_t=\frac{B_t}{\alpha_{t}}
\left[
\beta_{1}x - \log L_t(\alpha,-Z) + 
\sum_{k=1}^{t-1}\frac{\beta_{k+1}}{\alpha_{k}}\log L_k(\alpha, -Z)
\right], \quad t=2,\dots,T.
\end{align*}
\end{prop}

\begin{proof}
Let $t\ge 2$ (the case $t=1$ can be treated similarly). 
By Theorem \ref{thm:3.2} and Proposition \ref{prop:3.3}, 
the optimal intertemporal allocation $(X_t)$ of $x - Z$ satisfies
\begin{align*}
e^{-\alpha_{t}X_t/B_t}
&=M_t(\alpha, x-Z)
=L_t(\alpha, x-Z)\cdot\prod\nolimits_{k=1}^{t-1}
L_k(\alpha, x-Z)^{-\beta(k+1)/\alpha(k)}\\
&=e^{-\beta(t)x}\prod\nolimits_{k=1}^{t-1}
\left(e^{-\beta(k)x}\right)^{-\beta(k+1)/\alpha(k)}\\
&\qquad \times L_t(\alpha, -Z)\cdot\prod\nolimits_{k=1}^{t-1}
L_k(\alpha, -Z)^{-\beta(k+1)/\alpha(k)},
\end{align*}
whence
\begin{align*}
&\frac{\alpha_{t}}{B_t}X_t
=\left\{
\beta_{t}-\sum\nolimits_{k=1}^{t-1}\frac{\beta_{k}\beta_{k+1}}{\alpha_{k}}
\right\}x \\
&\qquad\qquad\qquad- \log L_t(\alpha, -Z) + 
\sum\nolimits_{k=1}^{t-1}\frac{\beta_{k+1}}{\alpha_{k}}\log L_k(\alpha, -Z).
\end{align*}
However, by simple calculation, we see that
\[
\beta_{t}-\sum\nolimits_{k=1}^{t-1}\frac{\beta_{k}\beta_{k+1}}{\alpha_{k}}
=\beta_{1}.
\]
Thus, the proposition follows.
\end{proof}


\section{Insurance pricing}\label{sec:4}

In this section, we apply the approach above 
to the computation of insurance premiums.


\subsection{Life insurance contract}\label{subsec:4.1}

We consider a life insurance contract with duration $T$ in which 
the insurer pays the insured $c_{t}$ dollars at time $t\in\T$ 
if the insured dies in the interval $(t-1,t]$. 
Here $c_{t}$'s are deterministic. The insured pays 
the insurer a one-time premium at time $t=0$.

We denote by $\tau$ the future life time of the insured, i.e., 
she/he dies at time $\tau$. 
We assume that $\tau$ is a random variable on $(\Omega,\mathcal{F},P)$ 
satisfying $\tau(\omega)>0$ for all $\omega\in\Omega$ and 
$P(\tau=t)=0$ for all $t\in [0,\infty)$. 

If the insured pays the insurer $H$ dollars as one time premium at 
time $t=0$, then the present value of the cashflow of the insurer 
is given by $H-Z$ with
\[
Z=\sum\nolimits_{t\in\T}\tilde c_{t}1_{(t-1<\tau\le t)},\qquad 
\tilde c_{t}:=c_{t}/B_t\quad\mbox{for $t\in\T$}.
\]

In the traditional pricing, 
the premium $H_0(Z)$ based on the {\itshape principle of equivalence} 
is often used: $H_0(Z)$ is defined by 
$E[H_{0}(Z)-Z]=0$ or $H_{0}(Z)=E[Z]$. 
If the interest rates are deterministic, 
$H_{0}(Z)$ is given by 
\[
H_{0}(Z)=\sum\nolimits_{t\in\T}\tilde c_{t}P(t-1<\tau\le t).
\]
Notice that this price lacks the {\itshape safety loading} 
if the real mortality table is used. 
Usually, insurance companies use modified mortality tables to 
ensure the necessary safety loading (see \S \ref{subsec:4.4} below).

We define a discrete-time process 
$(D_t)_{t\in\T}$ by
\[
D_t:=1_{(\tau\le t)},\qquad t=0,1,\dots,T.
\]
Then, $(D_t)_{t\in\T}$ is a $\{0,1\}$-valued nondecreasing process 
with $D_0=0$. 
Notice that for $t\in\T$, 
$D_t=0$ (resp., $D_t=1$) if and only if the insurer is alive (resp., 
dead) at time $t$. 
We denote by $(\mathcal{H}_t)_{t\in\T}$ the filtration associated with 
the process $(D_t)_{t\in\{0\}\cup\T}$:
\begin{equation}
\mathcal{H}_t:=\sigma(D_s: s=0,\dots,t),\qquad t=0,1,\dots,T.
\label{eq:eq:4.1}
\end{equation}

We consider the following conditional probabilities:
\begin{align*}
&q_{t}:=P(\tau\le t+1\ \vert\ \tau>t),\qquad t=0,\dots,T-1,\\
&p_{t}:=1-q_{t}=P(\tau> t+1\ \vert\ \tau>t),\qquad t=0,\dots,T-1.
\end{align*}
We have the following equalities:
\begin{equation*}
q_{t} + p_{t} = 1\quad (t=0,\dots,T-1),\quad 
q_{0}=P(\tau\le 1),\quad p_0=P(1<\tau).
\end{equation*}

We use the following well-known result.

\begin{lem}\label{lem:4.1}
The following assertions hold:
\begin{itemize}
\item[{\rm (a)}] $E[\left.D_t\right |\mathcal{H}_{t-1}]
=D_{t-1}+(1-D_{t-1})q_{t-1}$ for $t\in\T$.
\item[{\rm (b)}] $E[\left.(1-D_t)\right |\mathcal{H}_{t-1}]
=(1-D_{t-1})p_{t-1}$ for $t\in\T$.
\end{itemize}
\end{lem}


\subsection{Algorithm for the premium computation}\label{subsec:4.2}

The aim of this section is to derive an algorithm 
to compute the indifference premium of the life insurance contract. 
To this end, in addition to (EU), we assume the following conditions:
\begin{align*}
&\mbox{The interest rate process $(r_t)_{t\in\T}$ is deterministic.}
\tag{R}\\
&\mbox{The filtration $(\mathcal{F}_t)_{t\in\{0\}\cup\T}$ is given by 
$(\mathcal{H}_t)_{t\in\{0\}\cup\T}$ 
in (\ref{eq:eq:4.1}).}
\tag{F}
\end{align*}
The condition (R) implies that the riskless 
bond price process $(B_t)_{t\in\T}$ is also deterministic.

The $\sigma$-algebra $\mathcal{F}_T$ is generated by the 
followng decomposition of $\Omega$:
\[
\Omega
=(0<\tau\le 1)\cup(1<\tau\le 2)\cup\cdots\cup(T-1<\tau\le T)\cup (T<\tau).
\]
Hence, 
if $Z\in L^{\infty}(\Omega,\mathcal{F}_T,P)$, then $Z$ has 
the decomposition of the form 
\begin{equation*}
Z=\sum_{t=1}^{T}z_{t}1_{(t-1<\tau\le t)} +z_{T+1}1_{(T<\tau)}
\tag{Z}
\end{equation*}
with some real deterministic coefficients $z_{t}$, $t=1,\dots,T+1$. 
We also write $z(t)=z_t$. 
For example, in the life insurance contract considered in the previous 
section, we have $z_{t}=\tilde c_{t}$ for $t=1,\dots,T$ and $z_{T+1}=0$.

Recall $\beta_{t}$ from ($\beta$). 
For $Z\in L^{\infty}$ with 
representation (Z), 
we define the real deterministic sequence $(h_{t})_{t=0}^T$ 
by the following backward iteration:
\begin{equation*}
\begin{cases}
h_{T}=e^{z(T+1)},\\
h_{t-1}
=\left[e^{\beta(t)z(t)}q_{t-1} 
+ h_{t}^{\beta(t)}p_{t-1}\right]^{1/\beta(t)},\quad 
t=1.\dots,T.
\end{cases}
\tag{h}
\end{equation*}

Recall the definition of the process $(L_t(\alpha, -Z))_{t\in\T}$ from 
Section \ref{sec:2}.

\begin{prop}\label{prop:4.2}
We assume $({\rm EU})$, $({\rm R})$ and $({\rm F})$. Then, 
for $Z\in L^{\infty}$ with $({\rm Z})$, 
the process $(L_t(\alpha, -Z))_{t\in\T}$ is given by
\begin{equation*}
\begin{cases}
L_1(\alpha, -Z)=
e^{\beta(1)z(1)}D_1 + h_{1}^{\beta(1)}(1-D_1), & \\
L_t(\alpha, -Z)=
\exp\left[\beta_{t}\sum_{s=1}^{t-1}
\{z_{s} - z_{s+1}\}D_s\right] & \\
\qquad\qquad\qquad\qquad 
\times \left[e^{\beta(t)z(t)}D_{t} + h_{t}^{\beta(t)}(1-D_{t})\right], & 
t=2,\dots,T.
\end{cases}
\tag{L2}
\end{equation*}
\end{prop}

\begin{proof}
For simplicity, we write $L_t=L_t(\alpha, -Z)$. 
Since
\[
1_{(t-1<\tau\le t)}=D_{t}-D_{t-1}\quad(t=1,\dots,T),\quad 
1_{(T<\tau)}=1-D_T,
\]
we have
\begin{equation}
Z=\sum_{t=1}^{T-1}\{z_{t} - z_{t+1}\}D_t + z_{T}D_T+
z_{T+1}(1-D_T).
\label{eq:4.2}
\end{equation}
To prove (L2), we use backward mathematical induction with respect to $t$. 

First, if $t=T$, then from (\ref{eq:4.2}), 
\begin{equation*}
L_T=\exp\left[\beta_{T}
\sum\nolimits_{t=1}^{T-1}\{z_{t} - z_{t+1}\}D_t\right]
\cdot \exp\left[\beta_{T}\{z_{T}D_T + z_{T+1}(1-D_T)\}\right].
\end{equation*}
However, since $D_T$ is either $1$ or $0$ and 
$h_{T}=\exp(z_{T+1})$, we have
\[
\exp\left[\beta_{T}
\{z_{T}D_T + z_{T+1}(1-D_T)\}\right]
=e^{\beta(T)z(T)}D_T + h_{T}^{\beta(T)}(1-D_T),
\]
which implies (L2) with $t=T$.

Next, we assume that (L2) holds for $t\in\{2,\dots,T\}$. Then,
\begin{equation*}
E_{t-1}[L_t]
=\exp\left[
\beta_{t}
\sum\nolimits_{s=1}^{t-1} \{z_{s} - z_{s+1}\}D_s
\right]\cdot 
E_{t-1}\left[e^{\beta(t)z(t)}D_t + h_{t}^{\beta(t)}(1-D_t)\right],
\end{equation*}
where, as before, 
we write $E_t[X]$ for 
$E[X \vert \mathcal{F}_t]$. 
By Lemma \ref{lem:4.1},
\begin{align*}
&E_{t-1}\left[e^{\beta(t)z(t)}D_t + h_{t}^{\beta(t)}(1-D_t)\right]\\
&\qquad 
=e^{\beta(t)z(t)}\{D_{t-1} + (1-D_{t-1})q_{t-1}\} 
+ h_{t}^{\beta(t)}(1-D_{t-1})p_{t-1}\\
&\qquad 
=e^{\beta(t)z(t)}D_{t-1} 
+ \left[e^{\beta(t)z(t)}q_{t-1} + h_{t}^{\beta(t)}p_{t-1}\right]
(1-D_{t-1})\\
&\qquad 
=e^{\beta(t)z(t)}D_{t-1} 
+ h_{t-1}^{\beta(t)}(1-D_{t-1}).
\end{align*}
Hence, noting that $D_{t-1}$ is either $1$ or $0$, 
we obtain
\begin{align*}
L_{t-1}
&=E_{t-1}[L_t]^{\beta(t-1)/\beta(t)}\\
&=\exp\left[
\beta_{t-1}
\sum\nolimits_{s=1}^{t-1} \{z_{s} - z_{s+1}\}D_s
\right]\\
&\qquad\qquad\qquad\qquad 
\times 
\left[e^{\beta(t-1)z(t)}D_{t-1} 
+ h_{t-1}^{\beta(t-1)}(1-D_{t-1})\right]\\
&=\exp\left[\beta_{t-1}\sum\nolimits_{s=1}^{t-2}
\{z_{s} - z_{s+1}\}D_s\right]\\
&\qquad\qquad\qquad 
\times \left[e^{\beta(t-1)z(t-1)}D_{t-1} 
+ h_{t-1}^{\beta(t-1)}(1-D_{t-1})\right],
\end{align*}
which implies (L2) with $t-1$. Thus, (L2) holds for $t\ge 1$.
\end{proof}

We are ready to give the algorithms to compute the 
indifference premium $H(Z)$ and corresponding optimal allocation of 
the selling position $w+H(Z)-Z$. 
We see that the computations are reduced to those of 
$h_{t}$, $t=0,\dots,T$, in (h).

\begin{thm}\label{thm:4.3}
We assume $({\rm EU})$, $({\rm R})$ and $({\rm F})$. 
Let 
$Z\in L^{\infty}$ with 
representation $({\rm Z})$. 
Then, the following assertions hold.
\begin{itemize}
\item[{\rm (a)}] 
The indifference price $H(Z)$ is given by 
$H(Z)=\log h_{0}$.
\item[{\rm (b)}] 
Let $(X_t)\in\mathcal{A}(w+H(Z) - Z)$ be the optimal intertemporal 
allocation of $w+H(Z) - Z$: 
$\sum_{t\in\T}E[u_t(\tilde X_{t})]=U(w+H(Z) - Z)=U(w)$. 
Then, $(X_t)_{t\in\T}$ is given by
\begin{align*}
&X_1=\frac{B_1}{\alpha_{1}}
\left[
\beta_{1}(w+H(Z)) - 
\beta_{1}z_{1}\cdot 1_{(0<\tau\le 1)} 
- \beta_{1}\log h_{1}\cdot 1_{(1<\tau)}
\right],\\
&X_t
=\frac{B_t}{\alpha_{t}}
\left[
\beta_{1}(w+H(Z)) 
- \sum_{k=1}^t \beta_{k}z_{k}\cdot 1_{(k-1<\tau\le k)} 
-\beta_{t}\log h_{t}\cdot 1_{(t<\tau)}
\right.\\
&\left.\qquad\qquad\qquad\qquad 
+ \sum_{k=1}^{t-1}\frac{\beta_{k+1}}{\alpha_{k}}\beta_{k}\log h_{k}
\cdot 1_{(k<\tau)}
\right], \qquad
t=2,\dots,T. 
\end{align*}
\end{itemize}
\end{thm}

\begin{proof}
(a)\ Since $E[D_1]=q_{0}$ and $E[1-D_1]=p_{0}$, it follows from 
Proposition \ref{prop:4.2} that
\[
E[L_1(\alpha,-Z)]=e^{\beta(1)z(1)}q_{0} + h_{1}^{\beta(1)}p_{0}
=h_{0}^{\beta(1)}.
\]
The assertion (a) follows from this and Theorem \ref{thm:3.4} (c).

(b)\ 
From Proposition \ref{prop:4.2}, 
\begin{equation*}
\log L_1(\alpha, -Z)=
\beta_{1}z_{1}\cdot 1_{(0<\tau\le 1)} + \beta_{1}\log h_{1}\cdot 1_{(1<\tau)}
\end{equation*}
and, for $t=2,\dots,T$,
\begin{align*}
\log L_t(\alpha, -Z)&=
\beta_{t}\left[\sum\nolimits_{s=1}^{t-1}\{z_{s} - z_{s+1}\}\cdot D_s 
+ z_{t}\cdot D_{t}\right] \\
&\qquad\qquad\qquad\qquad\qquad\qquad\qquad\qquad 
 + \beta_{t}\log h_{t}\cdot (1-D_{t}),\\
&=\beta_{t}\left[\sum\nolimits_{s=1}^{t}z_{s}\cdot 1_{(s-1<\tau\le s)}\right] 
 + \beta_{t}\log h_{t}\cdot 1_{(t<\tau)}.
\end{align*}
We see that
\[
\beta_{t} - \sum_{k=s}^{t-1}\frac{\beta_{k+1}\beta_{k}}{\alpha_{k}}
=\beta_{s},\qquad 1\le s<t\le T.
\]
Hence, 
we have, for $t=2,\dots,T$,
\begin{align*}
& \log L_t(\alpha,-Z) - 
\sum_{k=1}^{t-1}\frac{\beta_{k+1}}{\alpha_{k}}\log L_k(\alpha, -Z)\\
&
=
\beta_{t}\log h_{t}\cdot 1_{(t<\tau)}
-\sum_{s=1}^{t-1}\frac{\beta_{s+1}}{\alpha_{s}}\beta_{s}\log h_s
\cdot 1_{(s<\tau)}
\\
&\quad 
+\beta_{t}\left[\sum\nolimits_{s=1}^{t}z_{s}\cdot 1_{(s-1<\tau\le s)}\right] 
-\sum_{k=1}^{t-1}\frac{\beta_{k+1}}{\alpha_{k}}
\beta_{k}\left[\sum\nolimits_{s=1}^{k}z_{s}\cdot 1_{(s-1<\tau\le s)}\right]\\
&=\beta_{t}\log h_{t}\cdot 1_{(t<\tau)}
-\sum_{s=1}^{t-1}\frac{\beta_{s+1}}{\alpha_{s}}\beta_{s}\log h_s
\cdot 1_{(s<\tau)}\\
&\qquad\qquad 
+\beta_{t}z_{t}\cdot 1_{(t-1<\tau\le t)}
+\sum_{s=1}^{t-1}
\left[\beta_{t} - \sum\nolimits_{k=s}^{t-1}\frac{\beta_{k+1}}{\alpha_{k}}
\beta_{k}\right]z_{s}\cdot 1_{(s-1<\tau\le s)}\\
&=\beta_{t}\log h_{t}\cdot 1_{(t<\tau)}
-\sum_{s=1}^{t-1}\frac{\beta_{s+1}}{\alpha_{s}}\beta_{s}\log h_s\cdot 
1_{(s<\tau)}+\sum_{s=1}^{t}\beta_{s}z_{s}\cdot 1_{(s-1<\tau\le s)}.
\end{align*}
This and Proposition \ref{prop:3.5} yield 
the assertion (b) with $t=2,\dots,T$. 
We can prove the case $t=1$ in the same way.
\end{proof}

\begin{rem}\label{rem:4.4}
In the premium calcluation method in Theorem \ref{thm:4.3}, 
we have assumed that the interest rate process $(r_t)_{t\in\T}$ is 
deterministic (the condition (R)). 
If instead we assume, e.g., that $(r_t)_{t\in\T}$ is a Markovian processs 
that is independent of $\tau$, then we obtain a similar 
pricing method that involves the transition probabilities of 
$(r_t)_{t\in\T}$. Such extensions to the case of random-interest-rate 
will be reported elsewhere.
\end{rem}


\subsection{Dependence on the risk aversion 
coefficients}\label{subsec:4.3}

As in the previous section, we assume (EU), (R) and (F). 
The aim of this section is to investigate the 
dependence of the indifference price $H(Z)$ 
on the absolute risk aversion coefficient set 
$\alpha=(\alpha_{1},\dots,\alpha_{t})\in (0,\infty)^{T}$. 
To emphasize the dependence on 
$\alpha$, we write $u_t(x;\alpha)$, $U_{\alpha}(Z)$, 
$H_{\alpha}(Z)$ and $h_t(\alpha)$ for the exponential utility function 
$u_t(x)$, utility $U(Z)$, indifference price $H(Z)$ and 
$h_{t}$ in (h), respectively. 
In what follows, $\alpha\to 0+$ (resp., $\alpha\to \infty$) means that 
$\alpha_{t}\to +0$ (resp., $\alpha_{t}\to \infty$) for all $t\in\T$.

To study the asymptotic behavior of $H_{\alpha}(Z)$
as $\alpha\to 0+$, we need the next lemma.

\begin{lem}\label{lem:4.5}
For $z\in \mathbb{R}$, $q\in [0,1]$, and 
$g:(0,\infty)\to (0,\infty)$ with limit 
$c:=\lim_{x\to 0+}\log g(x)\in \mathbb{R}$, we 
define $f(x):=\left[qe^{zx} + (1-q)g(x)^{x}\right]^{1/x}$ for 
$x>0$. 
Then,
\[
\lim_{x\to 0+}\log f(x)=qz+(1-q)c.
\label{eq:f-lim}
\]
\end{lem}

\begin{proof}
Take $\varepsilon>0$. If $x$ is positive and sufficiently close to $0$, then
\[
\frac{1}{x}\log \left[qe^{zx} + (1-q)e^{(c-\varepsilon)x}\right]
\le 
\log f(x)
\le 
\frac{1}{x}\log \left[qe^{zx} + (1-q)e^{(c+\varepsilon)x}\right],
\]
which yields
\[
qz+(1-q)(c-\varepsilon)
\le \liminf_{x\to 0+}\log f(x)
\le \limsup_{x\to 0+}\log f(x)\le qz+(1-q)(c+\varepsilon).
\]
Since $\varepsilon>0$ is arbitrary, the lemma follows.
\end{proof}

For $Z\in L^{\infty}$ with representation (Z), 
we have
\[
E[Z]=\sum_{t=1}^Tz_{t}P(t-1<\tau\le t) + z_{T+1}P(T<\tau).
\]
We define $H_{\infty}(Z)$ by
\[
H_{\infty}(Z):=\max\{z_{1},\dots,z_{T+1}\}.
\]
We can view $E[Z]$ (resp., $H_{\infty}(Z)$) as a lower (resp., upper) 
bound for any reasonable price of $Z$. 
From the next theorem, we see that $H_{\alpha}(Z)$ takes any value 
in $(E[Z], H_{\infty}(Z))$ by a suitable choice of $\alpha\in (0,\infty)^{T}$.

\begin{thm}\label{thm:4.6}
We assume $({\rm EU})$, $({\rm R})$ and $({\rm F})$. 
We also assume $0<q_{t}<1$ for all $t=0,\dots,T-1$. 
Then, for $Z\in L^{\infty}$, 
the following assertions hold:
\begin{itemize}
\item[{\rm (a)}] 
$\ds E[Z] \le H_{\alpha}(Z)\le H_{\infty}(Z)$ for all 
$\alpha\in (0,\infty)^{T}$.
\item[{\rm (b)}] 
$\ds \lim_{\alpha\to 0+}H_{\alpha}(Z)=E[Z]$.
\item[{\rm (c)}] 
$\ds \lim_{\alpha\to \infty}H_{\alpha}(Z)=H_{\infty}(Z)$.
\item[(d)] For every $\pi\in (E[Z], H_{\infty}(Z))$ and 
$\alpha=(\alpha_1,\dots,\alpha_T)\in(0,\infty)^T$, 
there exists $p\in (0,\infty)$ such that 
$\pi=H_{p\alpha}(Z)$, where $p\alpha:=(p\alpha_1,\dots,p\alpha_T)$.
\end{itemize}
\end{thm}

\begin{proof}
(a) By (\ref{eq:3.1}), we have $u_t(x;\alpha)\le x$. 
Hence, for $W\in L^{\infty}$,
\begin{align*}
U_{\alpha}(W)
&=\sup\left\{\sum\nolimits_{t=1}^{T}E[u_t(\tilde X_t,\alpha)]: 
(X_t)\in\mathcal{A}(W)\right\}\\
&\le \sup\left\{E\left[\sum\nolimits_{t=1}^{T}\tilde X_t\right]: 
(X_t)\in\mathcal{A}(W)\right\}=E[W],
\end{align*}
which implies 
$0=U_{\alpha}(H_{\alpha}(Z) - Z)\le E[H_{\alpha}(Z) - Z]$ 
or $E[Z]\le H_{\alpha}(Z)$. 

By (h), we have $h_T(\alpha)\le \exp[H_{\infty}(Z)]$. Moreover, if
$h_t(\alpha)\le \exp[H_{\infty}(Z)]$, then
\[
h_{t-1}(\alpha)\le \left[
q_{t-1}e^{\beta(t)H_{\infty}(Z)} + p_{t-1}e^{\beta(t)H_{\infty}(Z)}
\right]^{1/\beta(t)}
=e^{H_{\infty}(Z)}.
\]
Thus we finally see that $h_{\alpha}(0)\le \exp[H_{\infty}(Z)]$. 
This and Theorem \ref{thm:4.3} (a) give $H_{\alpha}(Z)\le H_{\infty}(Z)$.

(b) We have $\beta\to 0+$ as $\alpha\to 0+$. Hence, by 
applying Lemma \ref{lem:4.5} iterately to 
\[
h_{t-1}(\alpha)
=\left[e^{\beta(t)z(t)}q_{t-1} 
+ h_t(\alpha)^{\beta(t)}p_{t-1}\right]^{1/\beta(t)},\quad 
t=1.\dots,T,
\]
with $x=\beta_{t}$, $q=q_{t-1}$, $z=z_{t}$, and $g(x)=h_t(\alpha)$, 
we see the existence of the limits 
$h_{t}(0):=\lim_{\alpha\to 0+}h_t(\alpha)$, $t=0,\dots,T$, 
satisfying
\begin{equation*}
\begin{cases}
\log h_{T}(0)=z_{T+1}, & \\
\log h_{t-1}(0)=
q_{t-1}z_{t} + p_{t-1}\log h_{t}(0), & t=1,\dots,T.
\end{cases}
\end{equation*}
From this, we get
\begin{equation*}
\log h_{0}(0)
=q_{0}z_{1}+\sum\nolimits_{t=1}^{T-1}\left(\prod\nolimits_{s=0}^{t-1}
p_s\right)q_{t}z_{t+1}
+\left(\prod\nolimits_{s=0}^{T-1}p_s\right)z_{T+1}.
\end{equation*}
However, we have $q_{0}=P(0<\tau\le 1)$, 
\[
p_{0}q_1
=P(\tau>1)P(\tau\le 2\vert \tau>1)=P(1<\tau\le 2),
\]
and more generally, 
\[
\left(\prod\nolimits_{s=0}^{t-1}
p_s\right)q_{t}=P(t<\tau\le t+1),
\qquad t=1,\dots,T-1.
\]
We also have 
$\prod\nolimits_{s=0}^{T-1}p_s=P(T<\tau)$. 
Thus
\[
\log h_{0}(0)=\sum_{t=1}^Tz_{t}P(t-1<\tau\le t) + z_{T+1}P(T<\tau)
=E[Z]
\]
or
\[
\lim_{\alpha\to 0+} H_{\alpha}(Z)
=\lim_{\alpha\to 0+}\log h_{0}(\alpha)=E[Z].
\]

(c) Let $H_{\infty}(Z)=z_{t_0}$ with $t_0\in\{1,\dots,T+1\}$. 
If $t_0\ge 2$, then
\begin{equation*}
h_{t_0-1}(\alpha)
=\left[
q_{t_0-1}e^{\beta(t_0)H_{\infty}(Z)} 
+ p_{t_0-1}h_{t_0}(\alpha)^{\beta(t_0)}\right]^{1/\beta(t)}
\ge q_{t_0-1}^{1/\beta(t_0)}e^{H_{\infty}(Z)},
\end{equation*}
which, together with (h), gives
\begin{equation*}
h_{t_0-2}(\alpha)
\ge p_{t_0-2}^{1/\beta(t_0-1)}h_{t_0-1}(\alpha)
\ge p_{t_0-2}^{1/\beta(t_0-1)}q_{t_0-1}^{1/\beta(t_0)}e^{H_{\infty}(Z)}.
\end{equation*}
Repeating this argument, we finally obtain
\[
h_{0}(\alpha)\ge \left(\prod\nolimits_{s=0}^{t_0-2}p_s^{1/\beta(s+1)}\right)
q_{t_0-1}^{1/\beta(t_0)}e^{H_{\infty}(Z)}.
\]
Similarly, if $t_0=1$, then 
$h_{0}(\alpha)\ge q_{0}^{1/\beta(1)}e^{H_{\infty}(Z)}$. 
Therefore, since $\beta\to \infty$ as $\alpha\to \infty$, we obtain
\[
\liminf_{\alpha\to\infty} H_{\alpha}(Z)
=\liminf_{\alpha\to\infty}\log h_{0}(\alpha)\ge H_{\infty}(Z).
\]
However, $H_{\alpha}(Z)\le H_{\infty}(Z)$ by (a), so that 
$\lim_{\alpha\to\infty} H_{\alpha}(Z)=H_{\infty}(Z)$.

(d) By the construction in (h), 
$h_{0}(\alpha)$, whence $H_{\alpha}(Z)=\log h_{0}(\alpha)$, 
is continuous in $\alpha\in (0,\infty)^{T}$. 
Therefore, the assertion (d) follows from (a)--(c).
\end{proof}


\subsection{Numerical examples}\label{subsec:4.4}

We compare the indifference pricing method in
Theorem \ref{thm:4.3} with traditional ones by 
applying them to the following same insurance contract:
\begin{itemize}
 \item Type of insurance: term mortality insurance.
 \item Age at issue: 30 years old.
 \item Sex: male.
 \item Term of contract: from 1 year to 30 years.
 \item Loading of premium: excluded.
 \item Mortality rate: Standard Mortality Table 2007 for mortality insurance
(made by the Institute of Actuaries of Japan).
 \item Discount rate: 2\%.
 \item Payment method: annual payment.
 \item Sum assured: 1 (during the entire contract term).
\end{itemize}
By using the notation in the previous sections, the 
aggregate risk $Z$ of this contract becomes
\[
Z=\sum_{t=1}^{T}\frac{1}{(1 + 0.02)^t}1_{(t-1<\tau\le t)}.
\]
The traditional pricing methods that we use here are as follows:

\begin{itemize}
\item[(1)] Traditional method without risk loading:
\[
\mbox{The premium $\mathrm{TP1}(T)$}
=\sum_{t=1}^T \frac{1}{(1 + 0.02)^t}P(t-1<\tau\le t).
\]
\item[(2)] Traditional method with risk loading:
\[
\mbox{The premium $\mathrm{TP2}(T)$}=\sum_{t=1}^T \frac{1}{(1 + 0.02)^t}Q'_t,
\]
where 
$Q'_t:=Q_t + 0.01\times\{Q_t(1-Q_t)\}^{1/2}$ with $Q_t:=P(t-1<\tau\le t)$.
\end{itemize}

As above, we write TP1$(T)$ and TP2$(T)$ for the premiums of the 
contract with $T$ years of term obtained by 
the traditional pricing methods (1) and (2), respectively. 
For the values 
$a=1.0$, $1.5$, $2.0$ and $2.5$, we denote by $\mathrm{IP}_{a}(T)$ 
the premium of the same contract obtained by the indifference 
pricing method in Theorem \ref{thm:4.3} with $\alpha(t)\equiv a$. 
We also write $\mathrm{IP}_{\rm fit}(T)$ for the premium 
of the same contract calculated by the pricing method in Theorem \ref{thm:4.3} 
with $\alpha(t)=0.6+0.36\sqrt{t}$, the form of which 
is chosen to fit the graph of the indifference 
prices to that of TP2. We used the nonlinear least-squares 
to determine the form of $\alpha(t)$ for $\mathrm{IP}_{\rm fit}(T)$. 

In Figures 4.1--4.3, we plot the graphs of 
TP1, TP2, $\mathrm{IP}_{a}$, and $\mathrm{IP}_{\rm fit}$. 
We see that the fitted premiums $\mathrm{IP}_{\rm fit}(T)$ 
simultaneously approximate the corresponding traditional prices TP2$(T)$ 
well. 
We have repeated this procedure for various prices and 
obtained good fits in most cases. 
This observation suggests the following {\it implied utility
approach\/} to coherent pricing: insurance companies estimate
their implied utility functions by applying this method to 
existing products, and then refers to them in pricing other
products.

\begin{figure}[htbp]
\begin{center}
\includegraphics
[width=350pt,height=240pt]
{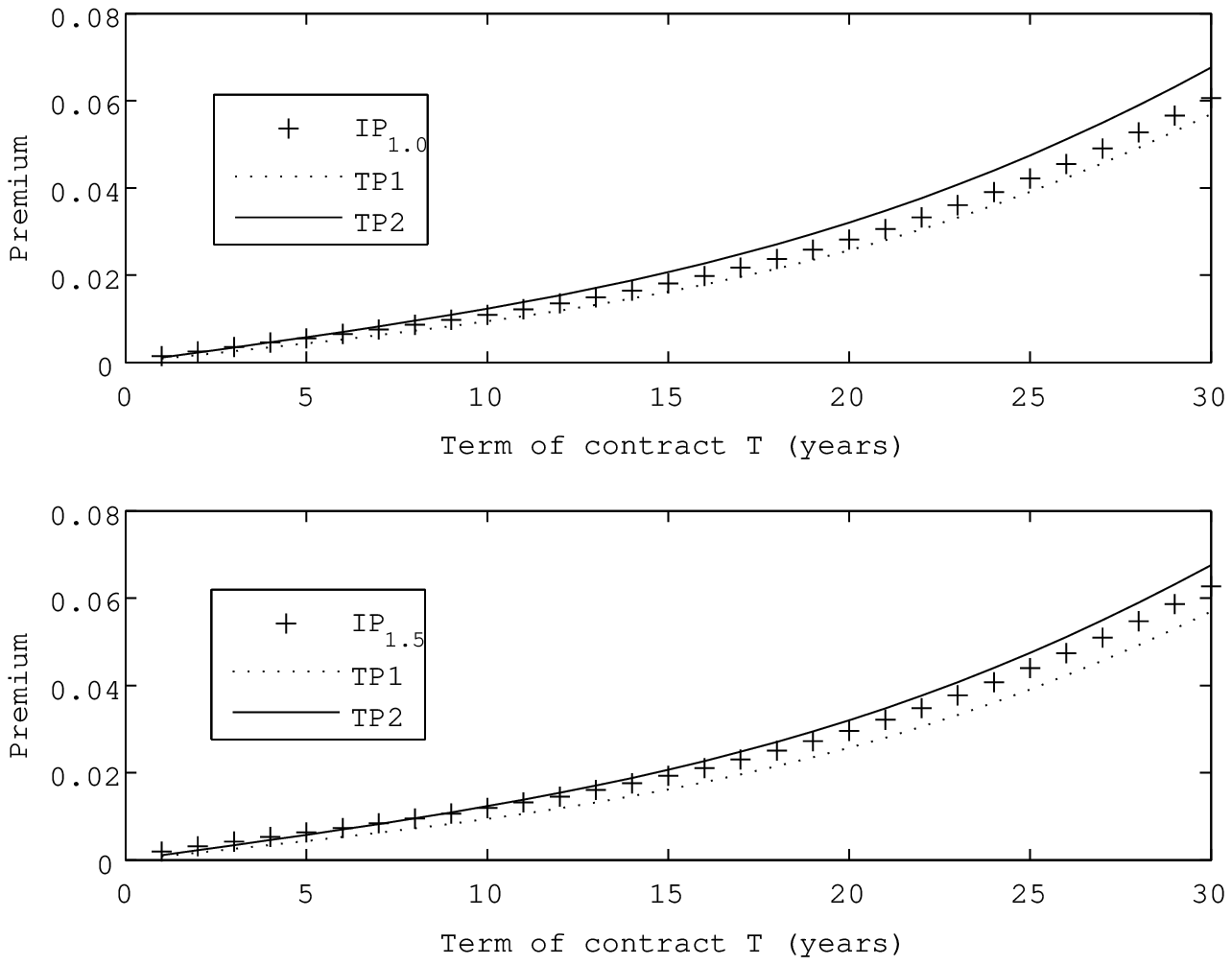}
\caption{TP1 and TP2  vs.\ 
$\mathrm{IP}_{1.0}$ and $\mathrm{IP}_{1.5}$.}
\end{center}
\label{fig:4.1}
\end{figure}

\begin{figure}[htbp]
\begin{center}
\includegraphics
[width=350pt,height=240pt]
{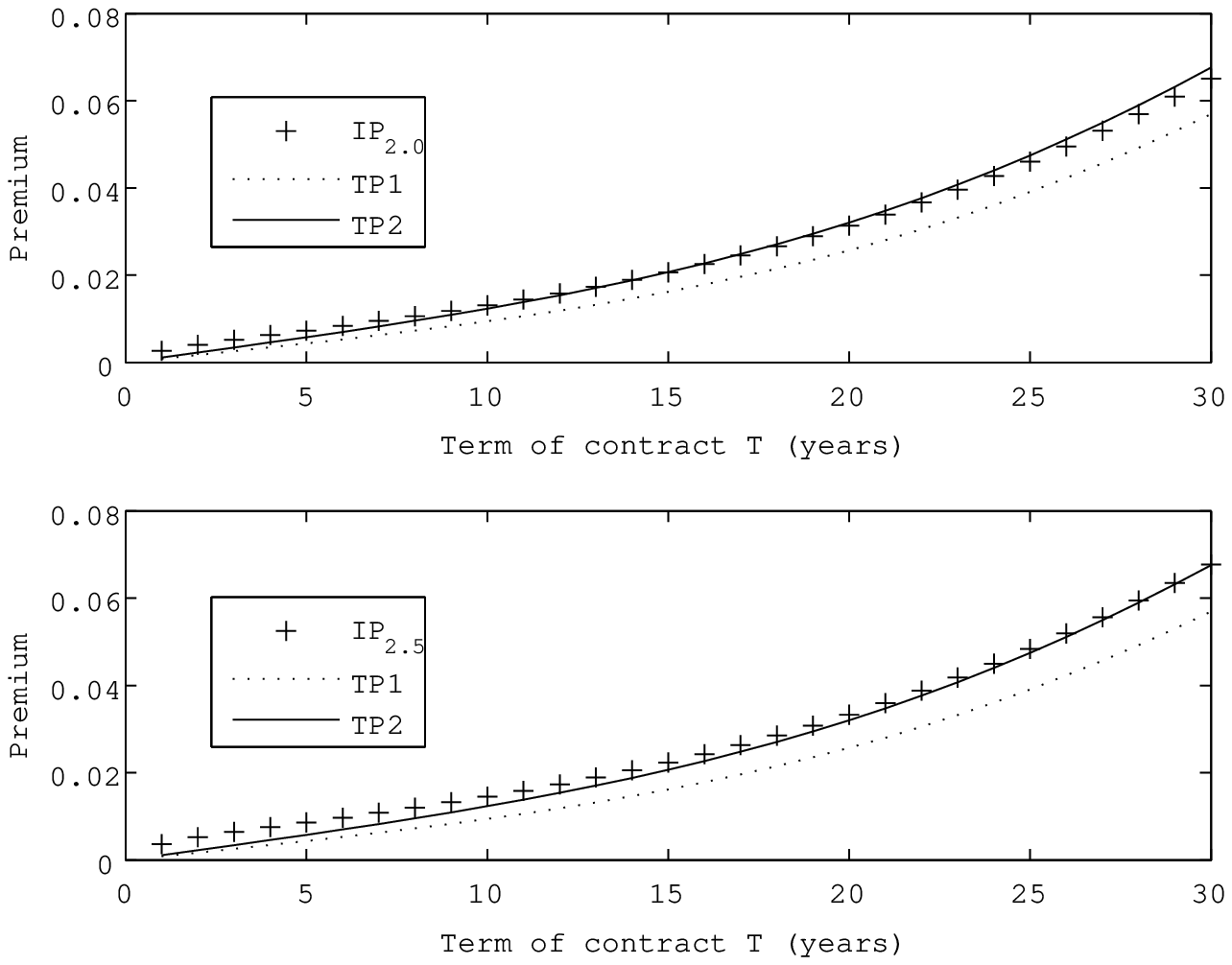}
\caption{TP1 and TP2  vs.\ 
$\mathrm{IP}_{2.0}$ and $\mathrm{IP}_{2.5}$.}
\end{center}
\label{fig:4.2}
\end{figure}

\begin{figure}[htbp]
\begin{center}
\includegraphics
[width=350pt,height=240pt]
{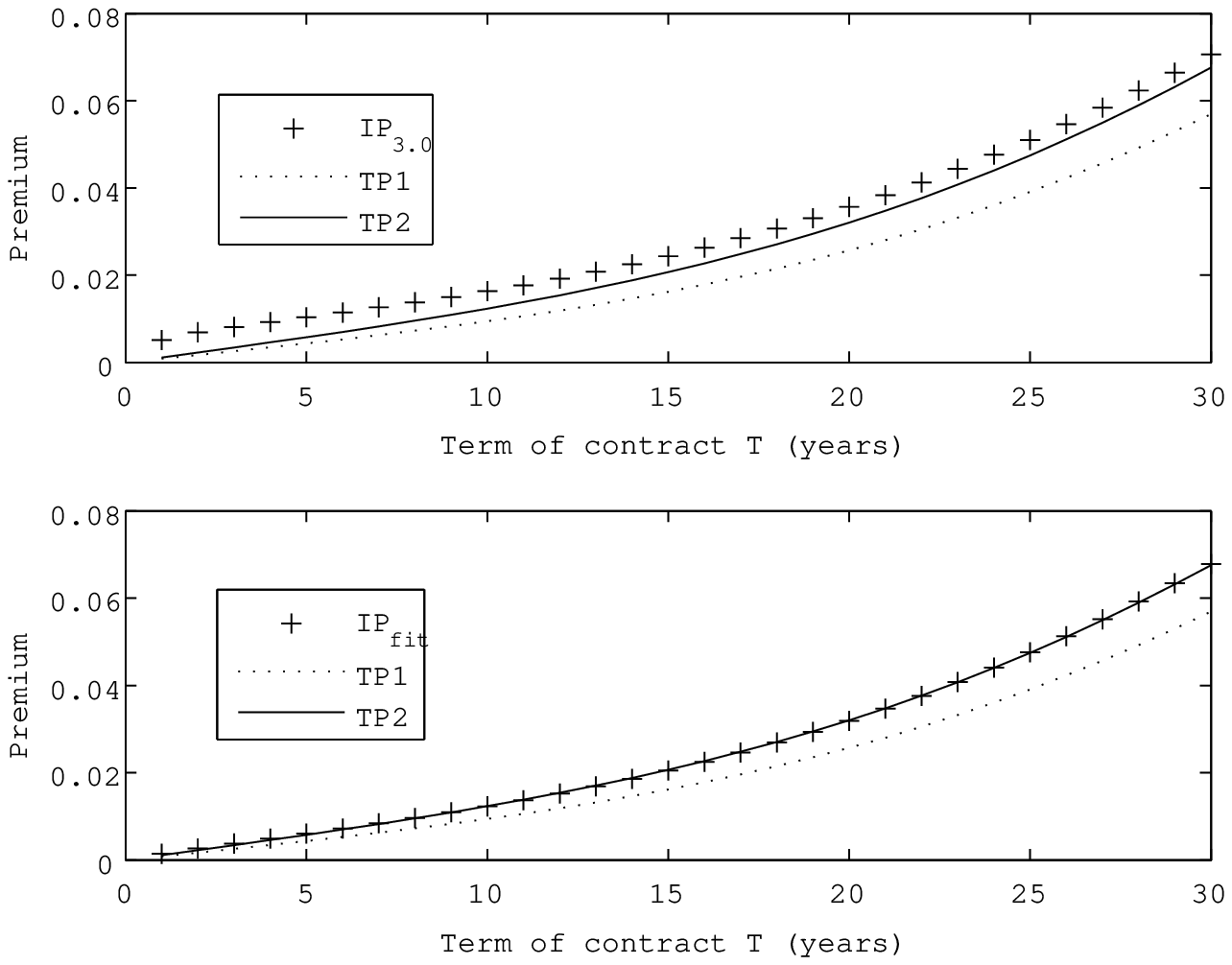}
\caption{TP1 and TP2  vs.\ 
$\mathrm{IP}_{3.0}$ and $\mathrm{IP}_{\rm fit}$.}
\end{center}
\label{fig:4.3}
\end{figure}


\end{document}